\documentclass[letterpaper, twocolumn, superscriptaddress,amsmath,amssymb,aps,prb]{revtex4-2} 

\usepackage[colorlinks=true]{hyperref}
\usepackage{graphicx}
\usepackage{dcolumn}
\usepackage{bm}
\usepackage{braket}

\DeclareMathOperator{\Tr}{Tr}

\begin{document}

\title{Information Scrambling and the Correspondence of Entanglement- and Operator Dynamics in Systems with Nonlocal Interactions} 
\author{Darvin Wanisch}
\email{darvin.wanisch@uni-jena.de}
\affiliation{Theoretisch-Physikalisches Institut, Friedrich-Schiller-Universit\"at Jena, Max-Wien-Platz 1, 07743 Jena, Germany}
\affiliation{Helmholtz-Institut Jena, Fr\"obelstieg 3, 07743 Jena, Germany}
\affiliation{GSI Helmholtzzentrum f\"ur Schwerionenforschung GmbH, Planckstra\ss e 1, 64291 Darmstadt, Germany}

\author{Juan Diego Arias Espinoza}
\email{j.d.ariasespinoza2@uva.nl}
\affiliation{Institute for Theoretical Physics, Institute of Physics, University of Amsterdam, Science Park 904, 1098 XH Amsterdam, Netherlands}

\author{Stephan Fritzsche}
\affiliation{Theoretisch-Physikalisches Institut, Friedrich-Schiller-Universit\"at Jena, Max-Wien-Platz 1, 07743 Jena, Germany}
\affiliation{Helmholtz-Institut Jena, Fr\"obelstieg 3, 07743 Jena, Germany}
\affiliation{GSI Helmholtzzentrum f\"ur Schwerionenforschung GmbH, Planckstra\ss e 1, 64291 Darmstadt, Germany}

\begin{abstract}
How fast quantum information scrambles such that it becomes inaccessible by local probes turns out to be central to various fields. Motivated by recent works on spin systems with nonlocal interactions, we study information scrambling in different variants of the Ising model. Our work reveals that nonlocal interactions can induce operator dynamics not precisely captured by out-of-time-order correlators (OTOCs). In particular, the operator size exhibits a slowdown in systems with generic powerlaw interactions despite a highly nonlinear lightcone. A recently proposed microscopic model for fast scrambling does not show this slowdown, which uncovers a distinct analogy between a local operator under unitary evolution and the entanglement entropy following a quantum quench. Our work gives new insights on scrambling properties of systems in reach of current quantum simulation platforms and complements results on possibly observing features of quantum gravity in the laboratory.
\end{abstract}

\maketitle

\textit{Introduction.---}
The dynamics of quantum information under unitary evolution lies at the heart of numerous ongoing questions in theoretical physics \cite{rev_dynqi_lewisswan,rev_neq_eisert}. Due to significant improvements on quantum simulation platforms \cite{sim_gases,sim_gases2,sim_ion,sim_rydberg,sim_rydberg2}, we are nowadays able to probe information dynamics of simple quantum lattice models in an experimental environment \cite{islam_measure_lab,kaufman_measure_lab,zoller_measure_1,zoller_measure_2,zoller_measure_lab,vermersch_otoc_measure}. In particular, in nonintegrable many-body systems, initial local quantum information can spread under unitary evolution such that local measurements are insufficient to reconstruct it at later times. This \textit{scrambling of quantum information} has received a great deal of attention lately. It is inherently related to thermalization \cite{thermalization_deutsch,thermalization_srednicki} and its absence \cite{mbl_review_serbyn,mbscars_review_serbyn}, as well as the simulability of many-body systems \cite{simulability_schuch}, and even quantum gravity \cite{gravity_qi}.  

\textit{Information Scrambling.---}
One particular probe of scrambling relates to the growing support and complexity of local operators under unitary evolution \cite{rev_otoc_swingle,altman_op_universival,krylov_compl_sonner}, known as \textit{operator spreading}. We can diagnose the spread of a local operator $\mathcal{W}$ via the \textit{squared commutator} with an auxiliary operator $\mathcal{V}$ at some distant site $r$
\begin{align}
    \label{otoc}
    C_r\left(t\right)=\frac{1}{2}\left\langle\left[\mathcal{W}\left(t\right),\mathcal{V}_{r}\right]^\dagger\left[\mathcal{W}\left(t\right),\mathcal{V}_{r}\right]\right\rangle\,,
\end{align}
where the expectation value is either evaluated in some pure state $\Ket{\Psi}$, or a thermal state $\rho_\mathrm{th}\sim e^{-\beta\mathcal{H}}$. Once the operator $\mathcal{W}\left(t\right)$ has spread such that its support overlaps with $\mathcal{V}$, Eq.\,\eqref{otoc} begins to grow and saturates afterwards at some finite value. By varying $r$, one can track how the operator spreads over the system's degrees of freedom. If $\mathcal{W}$ and $\mathcal{V}$ are unitary, the nontrivial part of Eq.\,\eqref{otoc} is determined by the OTOC, i.e., $F_r\left(t\right)=\left\langle \mathcal{W}\left(t\right)\mathcal{V}_{r} \mathcal{W}\left(t\right)\mathcal{V}_{r} \right\rangle$.

In a local quantum system, an emergent \textit{lightcone} constrains information propagation in accordance with the Lieb-Robinson bound \cite{lieb_robinson}.
However, today's experimental platforms often entail nonlocal (powerlaw) interactions $\sim1/r^\alpha$, leading to Hamiltonians that do not necessarily comply the assumption of locality. Seminal studies on systems with powerlaw interactions and experiments with trapped ions revealed vastly different nonequilibrium physics \cite{lr_spread_schachenmayer,lr_spread_hauke,eisert_break_locality,xy_spread_richerme,xy_spread_jurcevic}, e.g., \textit{the breakdown of quasilocality}. That is, information can propagate faster than allowed by the Lieb-Robinson bound. Since then, many works have focused on generalized bounds for systems with powerlaw interactions \cite{finite_speed_lr_lucas,lieb_robinson_imp_else,strictly_lightcones_kuwahara,guo_sig+scram,lc_powerlaw_new_lucas}, the akin process of operator spreading \cite{locality_luitz,lr_lightcone_levy_swingle,otoc_liebrobinson_luitz} and improved protocols for information processing tasks such as state transfer \cite{strictly_lightcones_kuwahara,lc_hierachy_lucas,optimal_transfer_pl_gorshkov}. In a nutshell, powerlaw interactions can induce an emergent nonlinear 'lightcone', i.e., information about a local operator can spread superballistically over the system's degrees of freedom.

Surprisingly, powerlaw interactions typically lead to a slowdown of entanglement growth \cite{lr_spread_schachenmayer}. It can be logarithmically slow for $\alpha<d$, where $d$ is the spatial dimension \cite{lr_entgrowth_lerose,lerose_semicl}. The \textit{entanglement entropy} of a region $A$ regarding a nonequilibrium state $\Ket{\Psi\left(t\right)}=e^{-i\mathcal{H}t}\Ket{\Psi_0}$ is given by the Von Neumann entropy of its reduced density matrix $S_A\left(\Ket{\Psi\left(t\right)}\right)=-\Tr\left[\rho_A\log\left(\rho_A\right)\right]$. Essentially, it probes how information about $A$ becomes inaccessible by measurements on $A$ due to entanglement with an increasing number of degrees of freedom and is thereby a complementary probe of scrambling. It appears then at a first sight operator spreading and entanglement growth have opposite behavior in these systems, which has also led to arguments saying that in general these two are not related, as both reflect properties of different spaces, i.e., state space and operator space \cite{tmi_pappalardi}.

\textit{Fast Scrambling.---}
\begin{figure*}
    \centering
    \includegraphics[width=0.32\textwidth]{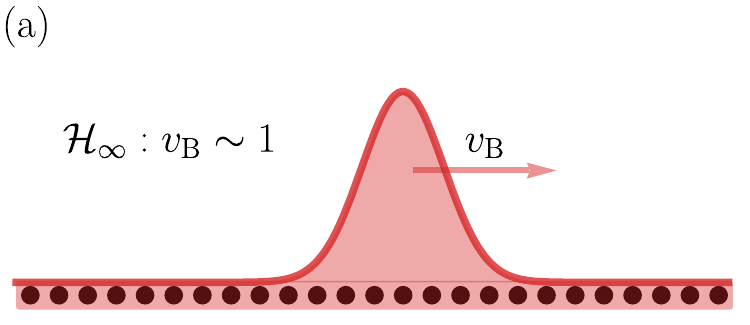}
    \includegraphics[width=0.32\textwidth]{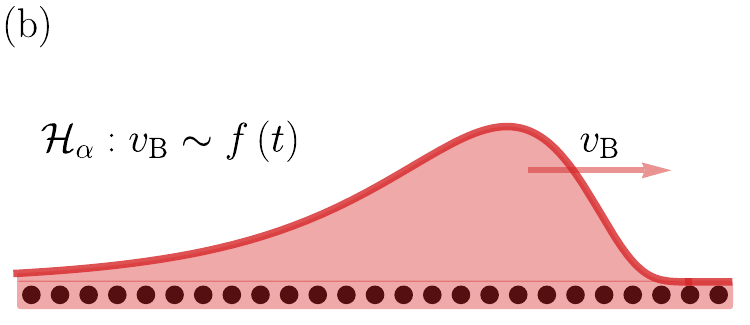}
    \includegraphics[width=0.32\textwidth]{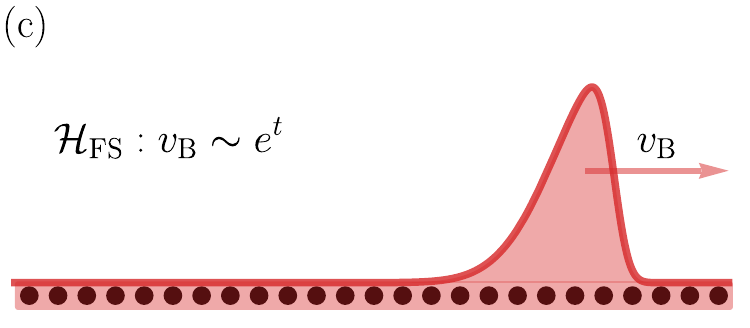}
    \caption{Illustration of operator dynamics regarding the considered Hamiltonians. Depicted is the temporal evolution of the operator density of an initially local operator. (a) The local Hamiltonian $\mathcal{H}_\infty$ is associated with a ballistic propagation of the operator front determined by a constant velocity $v_\mathrm{B}$. (b) For sufficiently small decay exponents $\alpha$, the powerlaw Hamiltonian $\mathcal{H}_\alpha$ is associated with a superballistic propagation of the operator front. Behind the operator front the operator density decays considerably slower compared to the local Hamiltonian. (c) Regarding the fast scrambling Hamiltonian $\mathcal{H}_\mathrm{FS}$, the operator front propagates exponentially fast with a rapid decay of the operator density behind the operator front.}
    \label{fig:pic_qid}
\end{figure*}
Recently, systems with nonlocal interactions appeared in connection to the correspondence of anti-de Sitter space and conformal field theories (AdS/CFT), where information scrambling has become a central topic \cite{bh_mirrors_hayden_preskill, fast_scramblers_sekino, towards_fast_scram_lashkari,bh_butterfly_schenker,hosur_chaos,bound_chaos_maldacena,ent_spread_mezei,op_growth_syk_roberts}. The property of \textit{fast scrambling}, i.e., a system with $N$ degrees of freedom, where $C_r\left(t\right)\sim O\left(1\right)\,\forall\, r$ in a time $t\sim\log\left(N\right)$ is believed to be characteristic of black holes \cite{fast_scramblers_sekino,towards_fast_scram_lashkari}, and holographic duals to theories of quantum gravity, e.g., the Sachdev-Ye-Kitaev (SYK) model \cite{sachdev_syk_magnet,kitaev_syk}. The highly complex structure of the latter, however, renders an experimental probe of fast scrambling challenging, and several proposals for simpler models with this property have appeared \cite{bentsen_fast_scrambling_coldatoms,belyansky_min_fast_scrambling,fast_scrambling_noholo_li,syk_simple_tezuka}. The proposals in Refs.\,\cite{belyansky_min_fast_scrambling,fast_scrambling_noholo_li} follow a similar structure: a fine tuned combination of a local Hamiltonian and a nonlocal all-to-all interaction $\sim 1/r^0$. Noteworthy, a slowdown of entanglement growth is absent in these models \cite{belyansky_min_fast_scrambling,fast_scrambling_noholo_li}. 

In addition, fast scrambling has been ruled out for systems with generic powerlaw interactions if $\alpha>d$, where lightcones are at most polynomial \cite{kuwahara_absence_fast_scrambling}. Whether or not this remains true for $\alpha\leq d$ is an open question. However, from an entanglement perspective, a system with (strong) powerlaw interactions starkly differs from, for example, the fast scrambling proposal in Ref.\,\cite{belyansky_min_fast_scrambling}.

If entanglement growth and operator spreading are both probes of information scrambling, is there a quantitative difference in operator dynamics in systems with unlike entanglement dynamics? In this work, we shed more light on this question and provide evidence for a distinct relation between entanglement growth and operator spreading. Our study is focused on strongly chaotic models. The dynamics of such systems is also of interest for studies on information dynamics in black holes \cite{bh_butterfly_schenker, chaos_bound_bh_Jahnke} and random unitary circuits \cite{nahum_hydro, keyserlingk_hydro,ru_circuits_correls_Nahum}, motivated by the fact that identifying universal properties of these extreme systems is a rich and important area of research.

Our main results are the following. First, we show that a local operator under unitary evolution exhibits, in a sense we make more precise, a growth that relates to the entanglement entropy following a quantum quench. This implies that operator dynamics in a fast scrambling model is in sharp contrast to that in systems with strong powerlaw interactions. 

Furthermore, we demonstrate the importance of alternative probes of operator dynamics, such as operator density and derived quantities, instead of the more common OTOCs. These other probes not only allow us to recognize known dynamical processes in systems with powerlaw interactions, such as an interplay of instantaneous growth of correlations combined with a slower local growth \cite{lr_spread_hauke,lr_kitaev_regemortel, exact_correl_LR_celovani}, but it also give us access to the structure of the operator within its support, which is not as evident from an analysis of the lightcone of the OTOC.

Finally, our results motivate us to extent the qualitative picture of operator dynamics in systems with local interactions \cite{keyserlingk_hydro,keyserlingk_hydro_conserv,nahum_hydro} to systems with nonlocal interactions. The latter can lead to widely different dynamics depending on their particular structure, which is pictorially illustrated in Fig.\,\ref{fig:pic_qid}.

\textit{Setup.---}
Let us consider the one-dimensional mixed-field Ising model of $N$ qubits with open boundary conditions
\begin{align}
    \label{ham_mfi}
    \mathcal{H}_\alpha=-\sum_{m<n}J_{mn}^\alpha\mathcal{Z}_m\mathcal{Z}_n-h_x\sum_m\mathcal{X}_m-h_z\sum_m\mathcal{Z}_m\,,
\end{align}
where $\mathcal{X}_m,\mathcal{Z}_m$ are Pauli $X,Z$ operators acting on site $m$. Interactions among the qubits decay with a powerlaw $J_{mn}^\alpha= J/\left\vert m-n\right\vert^\alpha$, where $J$ is the nearest-neighbor interaction strength, which we choose as our unit of energy, and we fix $h_x/J=-1.05$, $h_z/J=0.5$ throughout this work. Moreover, we consider the fast scrambling proposal from Ref.\,\cite{belyansky_min_fast_scrambling} whose Hamiltonian is given by

\begin{align}
    \label{ham_fs}
    \mathcal{H}_\mathrm{FS}=\mathcal{H}_{\infty}-\frac{1}{\sqrt{N}}\sum_{m<n}\mathcal{Z}_m\mathcal{Z}_n\,.
\end{align}

The local version of Eq.\,\eqref{ham_mfi}, i.e., $\mathcal{H}_{\infty}$ has been intensively studied in the context of information scrambling \cite{bh_butterfly_schenker,hosur_chaos,ent_spread_mezei,qi_speed_swingle,tmi_strong_th_sun} as it is strongly chaotic and exhibits \textit{strong thermalization} \cite{banuls_strong_th}.

We simulate the dynamics generated by Eq.\,\eqref{ham_mfi} and \eqref{ham_fs} numerically. Our methods consist of exact diagonalization (ED), a numerically exact method to obtain the action of the evolution operator on an initial state (EXPM) \cite{quspin}, and a matrix product state technique based on the time-dependent variational principle (TDVP) \cite{tdvp}. Regarding the quench dynamics, we consider an initial state with zero energy expectation value from the strong thermalization regime \cite{banuls_strong_th}. That is, local density matrices approach the infinite temperature ensemble, and the entanglement entropy saturates at a value expected for a Haar random state, i.e., the Page value $S_\mathrm{P}$ \cite{ent_rand_page}. In particular, we consider a fully polarized state along the $y$-direction $\ket{\Psi_0}=\ket{Y+}$, where $\ket{Y+}=\bigotimes_m\ket{y+}_m$. Our focus is on early to intermediate times as the considered Hamiltonians display substantially different dynamics there. The dynamics at late times turns out to be uniform, see Appendix\,\ref{app:lt_dyn}. We expect similar quench dynamics for different product states in $y$-direction, which all have zero energy expectation value.

\textit{Regimes.---}
In local quantum many-body systems, the entanglement entropy typically grows linear in time with an area-law growth rate, i.e., $S_A\left(\Ket{\Psi\left(t\right)}\right)\simeq \left\vert\partial A\right\vert v_\mathrm{E}t$, where $v_\mathrm{E}$ is the \textit{entanglement velocity}. Moreover, local operators are expected to spread ballistically with a characteristic velocity $v_\mathrm{B}$, known as the \textit{butterfly velocity}. Accordingly, $C_r\left(t\right)$ vanishes for $t\ll r/v_\mathrm{B}$, increases sharply around $t\sim  r/v_\mathrm{B}$, and saturates afterwards.

\begin{figure}
    \centering
    \includegraphics[width=1.0\linewidth]{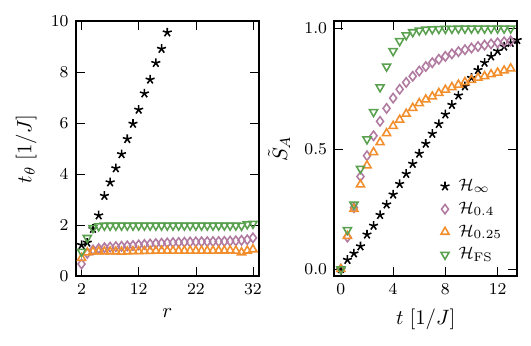}
    \caption{Quantum information scrambling for different Ising Hamiltonians. Local Hamiltonian $\mathcal{H}_{\infty}$, powerlaw Hamiltonian $\mathcal{H}_{\alpha}$ with $\alpha=0.4,0.25$ respectively, and fast scrambling Hamiltonian $\mathcal{H}_\mathrm{FS}$. Left panel: Spacetime contour $t_\theta$ determined by $C_r\left(t_\theta\right)=\theta$. Squared commutator is evaluated in the initial state $\Ket{Y+}$, $\mathcal{W},\mathcal{V}=\mathcal{Y}$, $\theta=0.5$, and $N=32$ (TDVP). Right panel: Half-chain entanglement entropy (normalized by the Page value, $\tilde{S}_{A} = S_A\left(\Ket{\Psi\left(t\right)}\right)/S_\mathrm{P}$) following a quench from $\Ket{Y+}$ for $N=26$ (EXPM).}
    \label{fig:ent_lc}
\end{figure}

A Hamiltonian like Eq.\,\eqref{ham_mfi} will generally possess a regime of $\alpha$ with effectively local dynamics \cite{lr_area_law_gong,strictly_lightcones_kuwahara,lc_powerlaw_new_lucas}, which we find to hold at least for $\alpha>2$. In this regime, the respective velocities of entanglement growth and operator spreading show a similar dependence on the exponent $\alpha$, see Appendix\,\ref{app:lin_reg}. This is the first evidence of a connection between entanglement growth and operator spreading, as they both diagnose a similar slowdown of information scrambling. For smaller exponents, the dynamics of $\mathcal{H}_\alpha$ becomes nonlocal. Here, we particularly focus on exponents $\alpha\leq1$, since fast scrambling might be possible in this regime. Moreover, a logarithmic ligthcone was proposed in Ref.\,\cite{lr_lightcone_levy_swingle} for $\alpha\leq 1/2$. Thus, in the following, we consider the local Hamiltonian $\mathcal{H}_\infty$, $\mathcal{H}_\alpha$ with $\alpha=0.4$, and $\alpha=0.25$ respectively, and $\mathcal{H}_\mathrm{FS}$ from Eq.\,\eqref{ham_fs}. 

\begin{figure}
    \centering
    \includegraphics[width=1.0\linewidth]{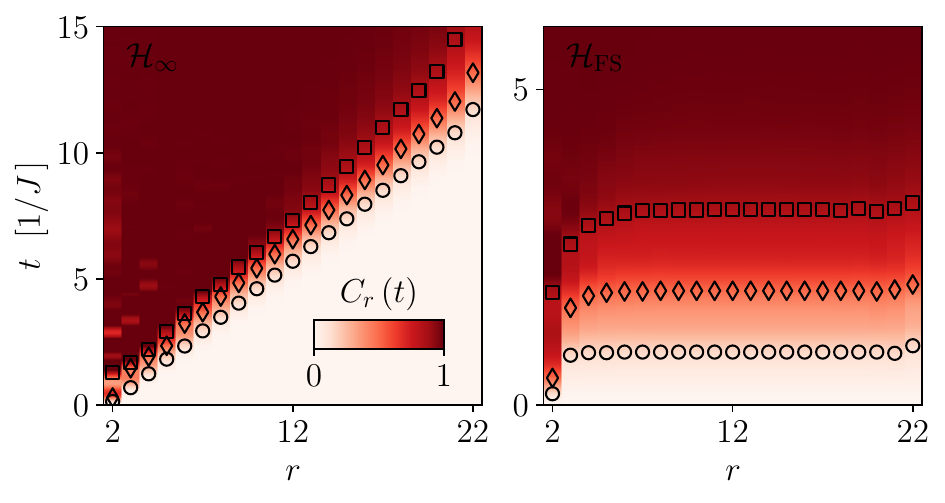}
    \caption{Spatiotemporal profile of the squared commutator $C_r\left(t\right)$, evaluated in the initial state $\ket{Y+}$ for $\mathcal{W},\mathcal{V}=\mathcal{Y}$. Markes show spacetime contours $t_\theta$ determined by $C_r\left(t_\theta\right)=\theta$ for $\theta=0.15,0.5,0.85$ respectively. Left panel refers to the local Hamiltonian $\mathcal{H}_\infty$ and right panel to the fast scrambler $\mathcal{H}_\mathrm{FS}$. System size is $N=22$ (EXPM).}
    \label{fig:lc_intro}
\end{figure}

To illustrate the dynamics for these Hamiltonians, we display the spacetime contour $t_\theta$ of the squared commutator in the left panel of Fig.\,\ref{fig:ent_lc}, where $t_\theta$ is defined by $C_r\left(t_\theta\right)=\theta$. We choose the Pauli $Y$ operator for $\mathcal{W}$, and $\mathcal{V}$, and evaluate $C_r\left(t\right)$ in the initial state $\Ket{\Psi_0}$. The left panel of Figure\,\ref{fig:ent_lc} shows the entanglement entropy following a quench from the initial state $\Ket{\Psi_0}$. As discussed earlier, both the system with powerlaw interactions and the fast scrambler induce a highly nonlinear lightcone as opposed to the linear one associated with the local system. Nevertheless, the slowdown of entropy growth that one observes for systems with powerlaw interactions is absent for the fast scrambler, hinting towards different dynamics, as we will discuss in more detail later. 

Let us further emphasize that the shape the lightcone, i.e., the dependence of $t_\theta$ on $r$, does not depend on the choice of $\theta$. This is demonstrated in Fig.\,\ref{fig:lc_intro}, which displays the spatiotemporal profile of $C_r\left(t\right)$ with several spacetime contours for the local Hamiltonian and the fast scrambler respectively.

\begin{figure}
    \centering
    \includegraphics[width=.75\linewidth]{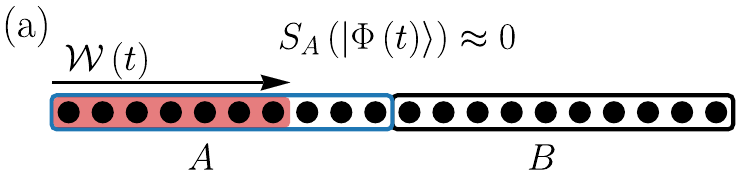}\\[12pt]
    \includegraphics[width=.75\linewidth]{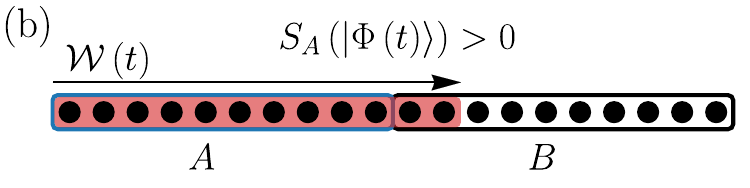}\\[12pt]
    \includegraphics[width=.75\linewidth]{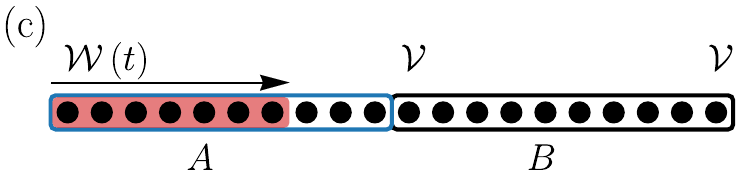}
    \caption{Entanglement growth of the operator state $\ket{\Phi\left(t\right)}$ from Eq.\,\eqref{phi_t}. (a) As long as the support of $\mathcal{W}\left(t\right)$ is confined to the region $A$, its entropy will vanish. (b) As soon as the support of $\mathcal{W}\left(t\right)$ exceeds $A$, $S_A\left(\Ket{\Phi\left(t\right)}\right)$ will deviate from zero since not all information about the operator is contained in $A$. (c) The specific setup we choose for Fig.\,\ref{fig:entropy_reverse}.}
    \label{fig:phi}
\end{figure}

\textit{Operator State.---}
To unveil the interplay between entanglement growth and operator spreading, it is constructive to consider states of the form 
\begin{align}
    \ket{\Phi\left(t\right)}:=\mathcal{W}\left(t\right)\ket{\Psi_0}\,.
\label{phi_t}
\end{align}
If $\Ket{\Psi_0}$ is a product state, the entropy $S_A\left(\Ket{\Phi\left(t\right)}\right)$ of a region $A$ that contains the initial position of $\mathcal{W}$ will vanish as long as the support of $\mathcal{W}\left(t\right)$ is confined to $A$, see Fig.\,\ref{fig:phi}(a). Once the operator has spread beyond $A$, entropy will grow as information about the operator is leaking out, see Fig.\,\ref{fig:phi}(b). Entanglement growth of Eq.\,\eqref{phi_t} is therefore in direct correspondence to the spread of $\mathcal{W}\left(t\right)$.

\begin{figure}
    \centering
    \includegraphics[width=1.0\linewidth]{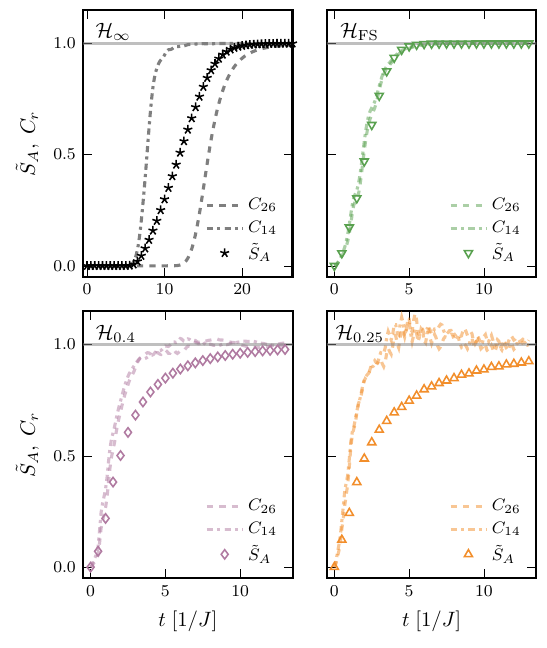}
    \caption{Entanglement growth, $\tilde{S}_A = S_A\left(\Ket{\Phi\left(t\right)}\right)/S_\mathrm{P}$ for $N = 26$ (EXPM), compared to the squared commutator $C_r\left(t\right)$, where the operator $\mathcal{W}$ is located at the leftmost site of the system, and $\mathcal{V}_r$ at the leftmost ($r=14$) or rightmost ($r=26$) site of the right block  $B$.}
    \label{fig:entropy_reverse}
\end{figure}

In Fig.\,\ref{fig:entropy_reverse}, we compare the entanglement entropy $S_A\left(\Ket{\Phi\left(t\right)}\right)$ of the left block $A$ with the squared commutator $C_r\left(t\right)$, where $r$ is chosen as either the leftmost or the rightmost site of the right block $B$, see Fig.\,\ref{fig:phi}\,(c) for an illustration. We chose the same initial state $\ket{\Psi_0}$ and operators $\mathcal{W},\mathcal{V}$ as in Fig.\,\ref{fig:ent_lc}. For the local model (upper left panel of Fig.\,\ref{fig:entropy_reverse}), entanglement growth agrees with the spatiotemporal structure of the squared commutator. That is, the entropy of the left block $A$ begins to grow, once the squared commutator diagnoses that the support of $\mathcal{W}\left(t\right)$ overlaps with the right block $B$. Shortly after the support has reached the rightmost site of $B$, entropy saturates in line with the squared commutator. The fast scrambler exhibits very different dynamics, i.e., highly nonlocal behavior as both entanglement entropy and squared commutator begin to grow immediately. However, both capture the same operator dynamics and behave similarly up to saturation, see the upper right panel in Fig.\,\ref{fig:entropy_reverse}.

On the contrary, this does not hold for powerlaw interactions. Although the initial entropy growth agrees with the squared commutator, we observe a slowdown  at intermediate times, similar to the ordinary quench scenario, see the lower panels in Fig.\,\ref{fig:entropy_reverse}. While the squared commutator grows rapidly and reaches its saturation value $C_r\left(t\right)\simeq 1$ (up to oscillations around it), the entropy $S_A\left(\Ket{\Phi\left(t\right)}\right)$ is still growing. Thus, information about $\mathcal{W}\left(t\right)$ is still leaking out of the left block $A$, although its support extends over the entire system for some time. This indicates that some part of the operator dynamics is not properly captured by the lightcone of squared commutator. Moreover, it suggests a slowdown of operator dynamics in the presence of (strong) powerlaw interactions similar to the entanglement entropy following a quantum quench.

\textit{Beyond the Quench.---}
So far, our focus was on the quench scenario, which is biased towards the (highly excited) initial state $\Ket{\Psi_0}$. This begs the question of how much of these insights are due to this choice. For a more general treatment, let us recall that any operator can be expanded in terms of a complete orthonormal operator basis, i.e., 
\begin{align}
    \mathcal{W}\left(t\right)=\sum_\Lambda c_\Lambda\left(t\right)\mathcal{S}_\Lambda\,,
    \label{op_strings}
\end{align}
where $\mathcal{S}_\Lambda=\bigotimes_{\lambda\in\Lambda}\mathcal{P}_\lambda$ are \textit{Pauli strings}, with $\mathcal{P}=\left\{\mathbf{1},\mathcal{X},\mathcal{Y},\mathcal{Z}\right\}$, and $\Tr\left(\mathcal{S}_\Lambda^\dagger\mathcal{S}_\Gamma^{}\right)/2^{N}=\delta_{\Lambda\Gamma}$. Considering an operator $\mathcal{W}$, initially supported on the leftmost site of the system, a useful measure based on the expansion \eqref{op_strings} is the \textit{operator density} \cite{loc_shocks_susskind,keyserlingk_hydro,nahum_hydro}
\begin{align}
    p_\ell\left(t\right)=\sum_{\left\vert\Lambda\right\vert=\ell} \left\vert c_\Lambda\left(t\right)\right\vert^2\,,
    \label{op_dens}
\end{align}
where the sum runs over all strings whose rightmost non-identity site is $\ell$. Note that $\sum_\ell p_\ell\left(t\right)=1\;\forall\;t$. Thus, Eq.\,\eqref{op_dens} measures how much weight of the operator is in strings whose support ranges from the first to the $\ell$-th site, i.e., strings of size $\ell$. The operator density is related to the squared commutator evaluated in the infinite temperature ensemble, see Appendix\,\ref{app:op_dens_alg}.

\begin{figure}
    \centering
        \includegraphics[width=1.0\linewidth]{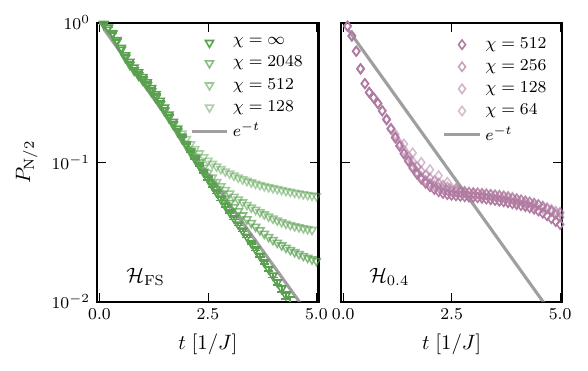}
    \caption{Decay of operator density in the left half of the system for $\mathcal{W}=\mathcal{Y}$, various bond dimensions $\chi$, and a system size of $N=24$ (TDVP). Left panel refers to the fast scrambler $\mathcal{H}_\mathrm{FS}$ and right panel to $\mathcal{H}_\alpha$ with $\alpha=0.4$. The value for $\chi = \infty$ for $\mathcal{H}_\mathrm{FS}$ was obtained by extrapolation. Error bars are smaller/similar than marker size}
    \label{fig:tdvp_op_dens}
\end{figure}

The slow entanglement growth regarding the operator state\,\eqref{phi_t} for $\mathcal{H}_\alpha$ suggests a slowdown of operator dynamics despite the superballistic propagation of the operator front, which is probed by the lightcone of the squared commutator. In the following, we investigate the decay of the operator density behind the operator front. In particular, we consider the total operator density in the left block $A$, i.e., 
$P_{N/2}=\sum_{\ell\leq N/2}p_\ell\left(t\right)$, which we expect to become exponentially small in the system size at late times \cite{bensa_circuit_twostep_otoc}. 

In Fig.\,\ref{fig:tdvp_op_dens}, we display the temporal evolution of $P_{N/2}$ for $\mathcal{W}=\mathcal{Y}$ and a system size of $N=24$ (TDVP), where we consider $\mathcal{H}_\alpha$ with $\alpha=0.4$, and $\mathcal{H}_\mathrm{FS}$ respectively. For the fast scrambler, $P_{N/2}$ approaches an exponential decay (diagonal line) with increasing bond dimension. For the powerlaw Hamiltonian, however, we observe a drastic slowdown of this decay, which remains upon increasing bond dimension. We observe a similar slowdown for other small values of $\alpha$, see Appendix\,\ref{app:tdvp}.

To summarize, we observe for the powerlaw system a superballistic propagation of the operator front together with a slower propagation of the tail of the operator, while for the fast scrambler the whole operator spreads exponentially fast. We can describe the dynamics of the former in terms of a fast process, leading to propagation of the operator front and the initial decay of operator density, and some remaining slow dynamics at which the tail of the operator propagates \footnote{In terms of quasiparticles \cite{exact_correl_LR_celovani, lr_spread_hauke, lr_kitaev_regemortel}, it is known that for $\alpha < d$, there are modes for which the group velocity diverges, and thus correlations spread instantaneously. Our observations confirm that picture, but importantly they make the connection to the dynamics of entanglement}. Together with our observations of the entanglement dynamics (Fig.\,\ref{fig:entropy_reverse}), this suggests that the operator front leads only to limited entanglement growth. This point is confirmed by the fact that $P_{N/2}$ can be recovered accurately with small bond dimension for the powerlaw system (Fig.\,\ref{fig:tdvp_op_dens}).

\begin{figure}
    \centering
    \includegraphics[width=1.0\linewidth]{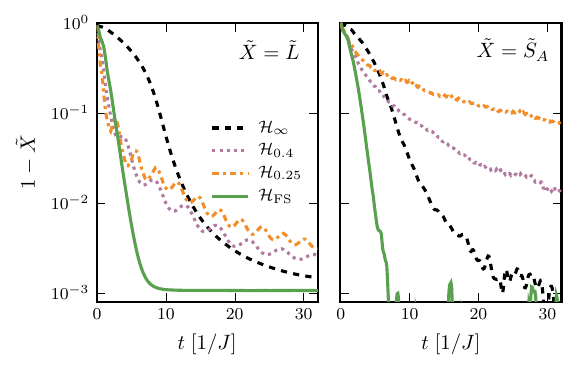}
    \caption{Left panel: Approach of operator size ($\tilde{L} = L\left[\mathcal{W}\left(t\right)\right]/L_\mathrm{Haar}$) towards its saturation value \eqref{op_size_lt}, where $\mathcal{W}=\mathcal{Y}$ and $N=16$ (ED). Right panel: Approach of half-chain entanglement entropy ($\tilde{S}_A = S_A\left(\Ket{\Psi\left(t\right)}\right)/S_\mathrm{P}$) towards the Page value, where $\ket{\Psi_0}=\ket{Y+}$ and $N=16$ (ED).}
    \label{fig:op_size}
\end{figure}

Let us further support this apparent slowdown of operator dynamics in the following. With use of Eq.\,\eqref{op_dens}, one can define the \textit{operator size} as \cite{loc_shocks_susskind,op_growth_syk_roberts,scrambling_alltoall_lucas}
\begin{align}
\label{op_extent}
    L\left[\mathcal{W}\left(t\right)\right]=\sum_\ell \ell\,p_\ell\left(t\right)\,.
\end{align}
Generally, one expects Eq.\,\eqref{op_extent} to grow monotonically and saturate at some value $\sim N$ at late times. For random unitary dynamics, the coefficients in \eqref{op_strings} should be uniformly distributed (excluding the identity) \cite{bensa_circuit_twostep_otoc}. Therefore, the operator density $p_\ell$ is on average determined by the number of strings with size $\ell$, i.e., $p_\ell\simeq 3\cdot 4^{\ell-1}/\left(4^N-1\right)$. The operator size\,\eqref{op_extent} under random unitary dynamics then becomes
\begin{align}
\label{op_size_lt}    
    L_\mathrm{Haar}=N\left(1+\frac{1}{4^N-1}\right)-\frac{1}{3}\approx N-\frac{1}{3}\,.
\end{align}
Equation \eqref{op_size_lt} is the average operator size of a random unitary of $N$ qubits drawn from the Haar measure.

To evaluate Eq.\,\eqref{op_extent} we compute the dynamics of $\mathcal{W}\left(t\right)$ using exact diagonalization (ED). We present the results of this calculation in the left panel of Fig.\,\ref{fig:op_size} for the same Hamiltonians as in Fig.\,\ref{fig:ent_lc} and Fig.\,\ref{fig:phi}, where we choose $\mathcal{W}=\mathcal{Y}$. In all the cases, the value of the operator size approaches $L_\textrm{Haar}$ from Eq.\,\eqref{op_size_lt} at late times. At short to intermediate times, we observe a clear analogy between the operator size and the entanglement entropy following a quench, where the latter is shown in the right panel of Fig\,\ref{fig:op_size} for the same system size. In particular, the operator size exhibits a slowdown for the powerlaw Hamiltonian $\mathcal{H}_\alpha$. In agreement with the observation in Fig.\,\ref{fig:tdvp_op_dens}, this slowdown is absent for the fast scrambler $\mathcal{H}_\mathrm{FS}$. Through this analogy, we are establishing a nontrivial correspondence between the dynamics in state- and operator space, through the lens of entanglement entropy and operator density/size. This correspondence is otherwise not evident through the conventional study of the lightcone of the squared commutator.

Qualitatively, we can understand the observed dynamics as follows. If we consider the information in our system to be initially encoded in our operator, this information starts to leak out of a region $A$ when the operator front (defined by the maximum of the operator density at a given time) crosses its boundary, and continues as the rest of the operator density exists the region. For the powerlaw system, the decay of the operator density behind the operator front is much slower compared to the fast scrambler. The total operator density in a region $A$, which includes the initial position of $\mathcal{W}$, i.e., $P_{A}$, therefore, remains large for a longer time.  In other words, information that is initially confined to $A$ leaks out much slower for the powerlaw Hamiltonian, which is manifested in a slowdown of entanglement growth. This qualitative picture of different classes of operator dynamics is summarized in Fig.\,\ref{fig:pic_qid}.

\textit{Conclusions and Outlook.---}
We found a connection between entanglement growth and operator spreading that reveals distinct classes of operator dynamics in the presence of nonlocal interactions. These classes are not clearly distinguishable by the squared commutator alone, at least not for system sizes of current numerical or experimental reach. In particular, the slowdown of entanglement entropy in systems with strong powerlaw interactions manifests in a slower decay of the operator density $p_\ell$ behind the operator front. Since $\sum_\ell p_\ell =1$ holds, a generally slower than exponential decay of $p_\ell$ may eventually slow down the operator front and thereby prohibit fast scrambling in systems with strong powerlaw interactions for large enough $N$. In addition, this behavior is in sharp contrast to the fast scrambler from Eq.\,\eqref{ham_fs}, which shows no slowdown in both operator size/density and entanglement entropy. 

Furthermore, this connection indicates that fast scrambling might be associated with universal entanglement dynamics. A recent study showed that fast scrambling is prohibited in models with a generic all-to-all term with prefactor $\sim 1/N^\gamma$ if $\gamma> 1/2$ \cite{scrambling_alltoall_lucas}. Moreover, the authors of Ref.\,\cite{belyansky_min_fast_scrambling} argued that for the Hamiltonian\,\eqref{ham_fs}, fast scrambling only occurs if $\gamma=1/2$. Interestingly, by further decreasing $\gamma$ from $1/2$, we observe a slowdown of entanglement growth, similar to our findings for powerlaw interactions. Future theoretical work may explore the relationship between entanglement growth and fast scrambling in microscopic quantum systems. 

An extension of this work may consider holographic models, which obey monogamy of mutual information \cite{tmi_holo_hayden}. The latter sets further restrictions on entanglement growth and is violated in systems with strong powerlaw interactions \cite{tmi_wanisch}. A refined understanding of entanglement growth in the presence of nonlocal interactions may result in explicit probes for holographic quantum matter.

One might also investigate the observed slowdown of operator dynamics in connection to prethermalization in systems with powerlaw interactions, e.g., in ion traps \cite{neyenhuis_pretherm}. 

Generally, the precise relationship between entanglement growth and operator spreading characterizes various nonequilibrium phenomena. To the best of our knowledge, there is no example where entanglement growth does not serve as a bottleneck of information dynamics, for example, linear entanglement growth but a superlinear lightcone. A throughout understanding of this relationship may improve our ability to probe nonequilibrium phenomena and phases of quantum matter.      

\section*{Acknowledgements}
We thank M. Peschke for sharing the framework for the TDVP calculations and for discussions. D. W. gratefully acknowledges support support from the Helmholtz Institute Jena and the Research School of Advanced Photon Science of Germany. J.D.A.E gratefully acknowledges support from an European Research Grant No. 677061 and from the Institute of Physics of the University of Amsterdam. The computational experiments were partly performed on resources of Friedrich Schiller University Jena supported in part by DFG grants INST 275/334-1 FUGG and INST 275/363-1 FUGG.

\appendix

\section{Late-Time Dynamics}
\label{app:lt_dyn}
\begin{figure}[h]
    \centering
    \includegraphics[width=.49\linewidth]{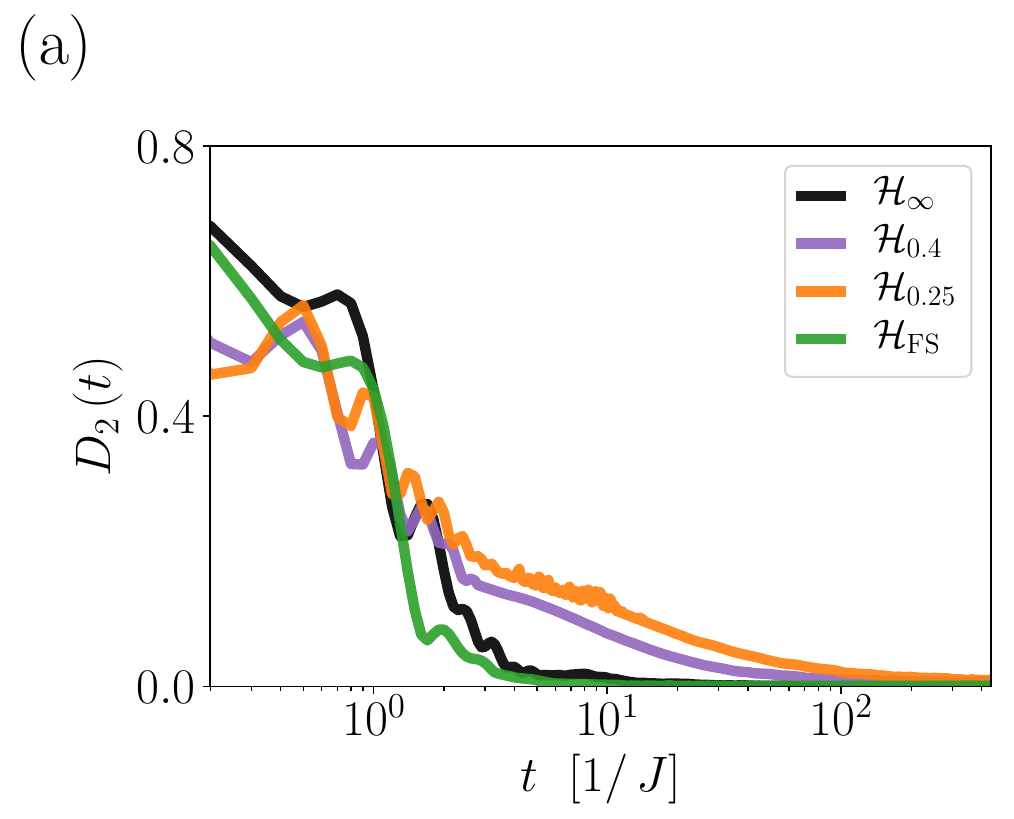}
    \includegraphics[width=.49\linewidth]{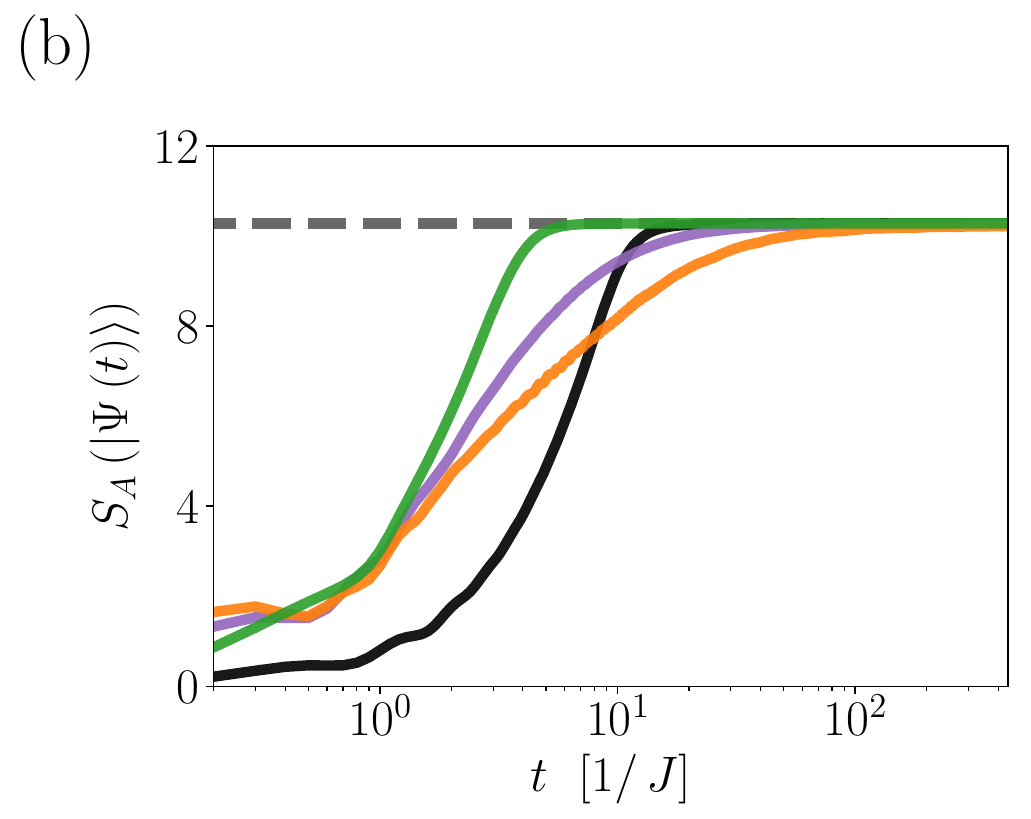}
    \caption{Late-time quench dynamics. (a) Distance of two-qubit density matrix to maximally mixed state. (b) Half-chain entanglement entropy. Gray dashed line shows the Page value $S_\mathrm{P}$. System size is $N=22$ (EXPM).}
    \label{fig:lt_dyn}
\end{figure}

In the following, we provide further details on the late-time dynamics of the considered Hamiltonians, regarding a quantum quench with initial state $\ket{Y+}$. Figure\,\ref{fig:lt_dyn}\,(a) shows the distance between the maximally mixed state and the reduced density matrix regarding two neighboring qubits in the middle of the system. We define this distance as the operator norm of the difference of the two density matrices. For all considered Hamiltonians, the reduced density matrix approaches the maximally mixed state at late times. For small values of $\alpha$, the powerlaw Hamiltonian $\mathcal{H}_\alpha$ is associated with a significant slowdown of this approach. Note that we observe similar behavior for other local density matrices. Furthermore, we display the half-chain entanglement entropy in Fig.\,\ref{fig:lt_dyn}\,(b). In all cases, the entanglement entropy approaches the Page value $S_\mathrm{P}$ at late times, see the dashed line. Summarizing, in the considered quench scenario, the late-time behaviour is the same for the different Hamiltonians that we studied. That is, local observables are determined by the expectation value in the infinite temperature ensemble and the entanglement entropy saturates at the Page value. 

\section{The Local Regime}
\label{app:lin_reg}

\begin{figure}[h]
    \centering
    \includegraphics[width=.5\linewidth]{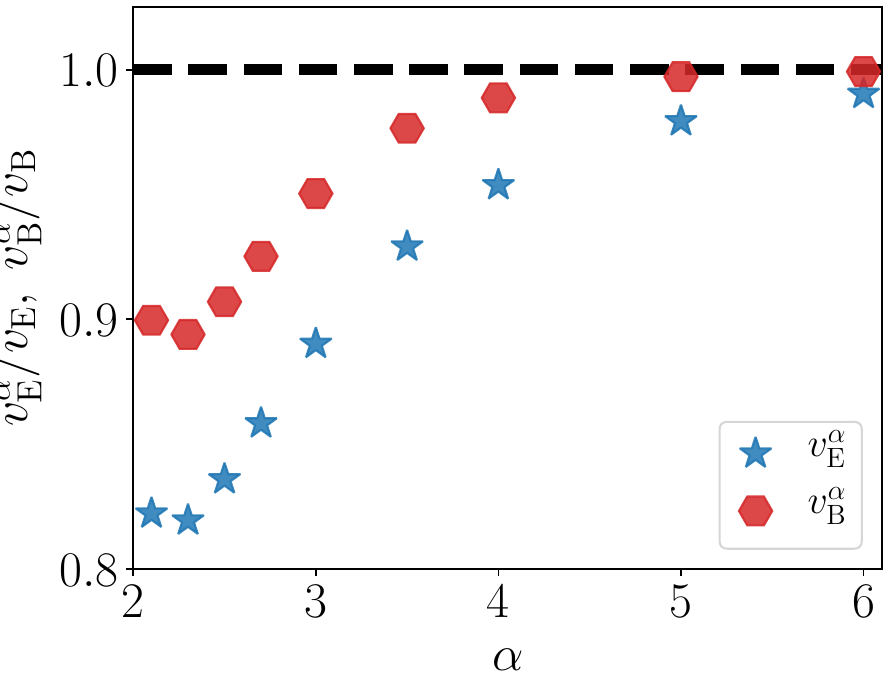}
    \caption{Velocities $v_\mathrm{E}^\alpha$ and $v_\mathrm{B}^\alpha$ as a function of the exponent $\alpha$, where $\Ket{\Psi_0}=\Ket{Y+}$, $\mathcal{W},\mathcal{V}=\mathcal{Y}$. System size is $N=24$ (EXPM). Dashed line indicates the value for the local Hamiltonian, i.e., $\mathcal{H}_\infty$.}
    \label{fig:vel}
\end{figure}

We have primarily focused our analysis on small values of the decay exponent $\alpha$ since the slowdown of entanglement growth and operator dynamics is most dominant in this case. As mentioned in the main text, there generally exists a regime of $\alpha$ with effectively local dynamics. Accordingly, in this regime, entanglement entropy exhibits a linear growth and operator spreading is bounded by a linear lightcone. Although we cannot rigorously prove where the transition to this regime occurs, for the model at hand our numerical results indicate that at least for $\alpha>2$, the dynamics are effectively local. As mentioned in the main text, the respective velocities $v_\mathrm{E}^\alpha$ and $v_\mathrm{B}^\alpha$ are similar renormalized in this regime, see Fig.\,\ref{fig:vel}. Hence, in this \textit{local regime} a connection between entanglement growth and operator spreading can be observed already, as they both diagnose a likewise slowdown of information scrambling. Moreover, this has experimentally relevance as the local dynamics of $\mathcal{H}_\alpha$ is accessible for a broader range of $\alpha$ on experimental platforms such as trapped ions, which are typically limited to $0\leq\alpha\leq 3$. We note that finite size effects on the velocities are negligible for the considered system sizes and conclude that the calculated velocities are universal properties of the Hamiltonian $\mathcal{H}_\alpha$ and the initial state $\ket{Y+}$ for all $N$.

\begin{figure}[h]
    \centering
    \includegraphics[width=.49\linewidth]{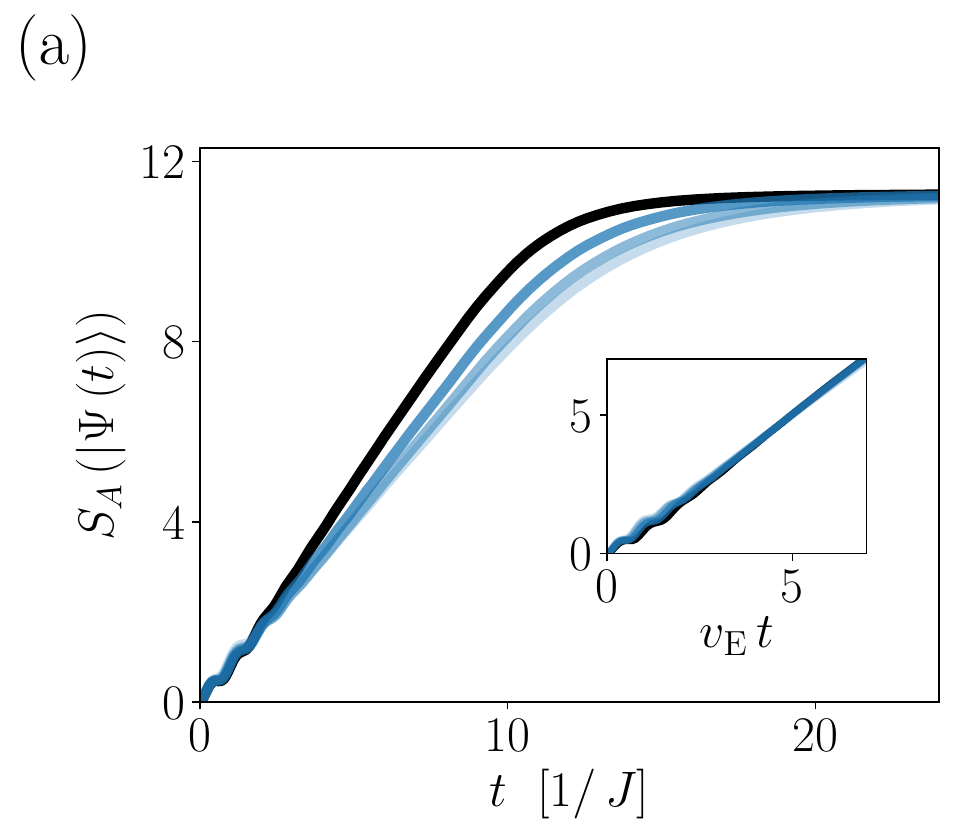}
    \includegraphics[width=.49\linewidth]{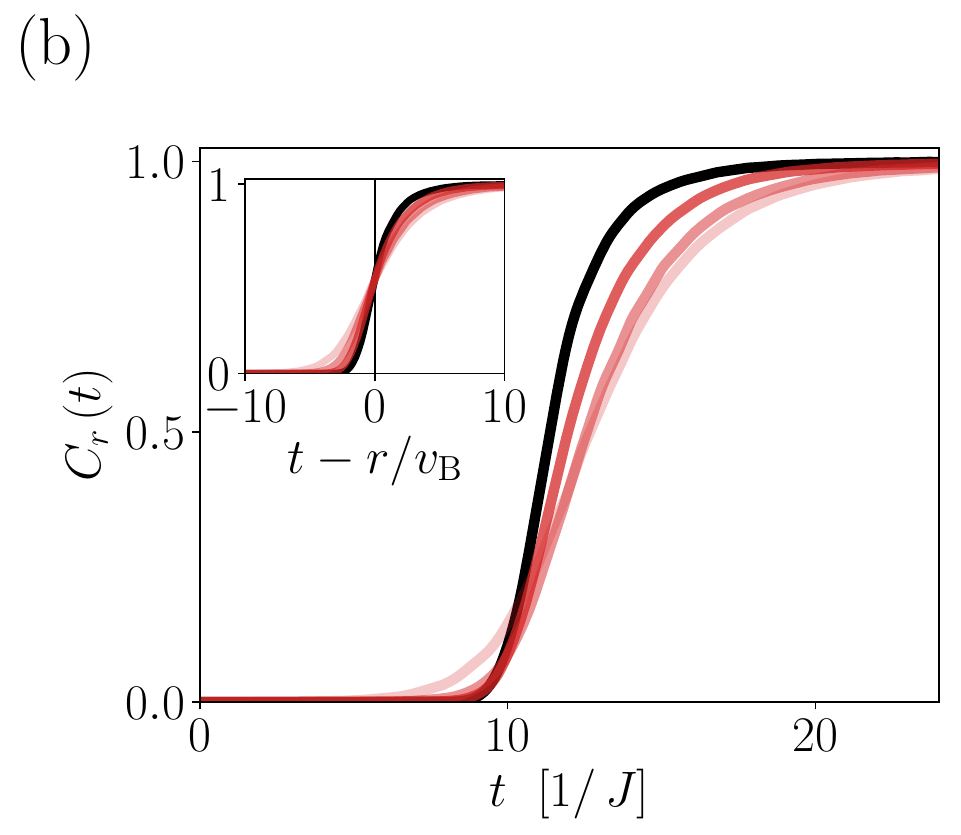}
    \caption{Entanglement growth and operator spreading for various values of $\alpha=\{\infty,3.0,2.5,2.1\}$. (a) Half-chain entanglement entropy following a quench with initial state $\ket{Y+}$, inset shows the collapse of all curves by rescaling time with the respective entanglement velocity $v_\mathrm{E}^\alpha$. (b) Squared commutator $C_r\left(t\right)$ evaluated in the initial state $\ket{Y+}$ for $r=20$. The inset shows the collapse at the operator front if time is rescaled according to the butterfly velocity $v_\mathrm{B}^\alpha$. Darker colors indicate larger values $\alpha$. System size is $N=24$ (EXPM)}
    \label{fig:entropy_norm}
\end{figure}

The effective local dynamics for $\alpha>2$ is further demonstrated in Fig.\,\ref{fig:entropy_norm}. In Fig.\,\ref{fig:entropy_norm}\,(a), the half-chain entanglement entropy for various values of $\alpha$ within the local regime is shown. With decreasing $\alpha$, the growth rate of entanglement entropy also decreases. The inset shows a clear collapse of the data if time is rescaled with the respective entanglement velocity $v_\mathrm{E}^\alpha$, which highlights the local dynamics in this regime. Figure\,\ref{fig:entropy_norm}\,(b) shows the squared commutator for various values of $\alpha$ within the local regime. In a similar vein, the growth of the squared commutator decreases with smaller $\alpha$. A collapse at the operator front, i.e., $C_r\left(t=r/v_\mathrm{B}^\alpha\right)$ can be observed if time is rescaled my means of the butterfly velocity, see the inset of Fig.\,\ref{fig:entropy_norm}\,(b). Moreover, we observe an increased broadening of the operator front with decreasing $\alpha$, which may be a first signature of the slower decay of the operator density behind hte operator front as discussed in the main text.

\section{Additional Data for Operator Size}
\label{app:op_size}

\begin{figure}[h]
    \centering
    \includegraphics[width=.98\linewidth]{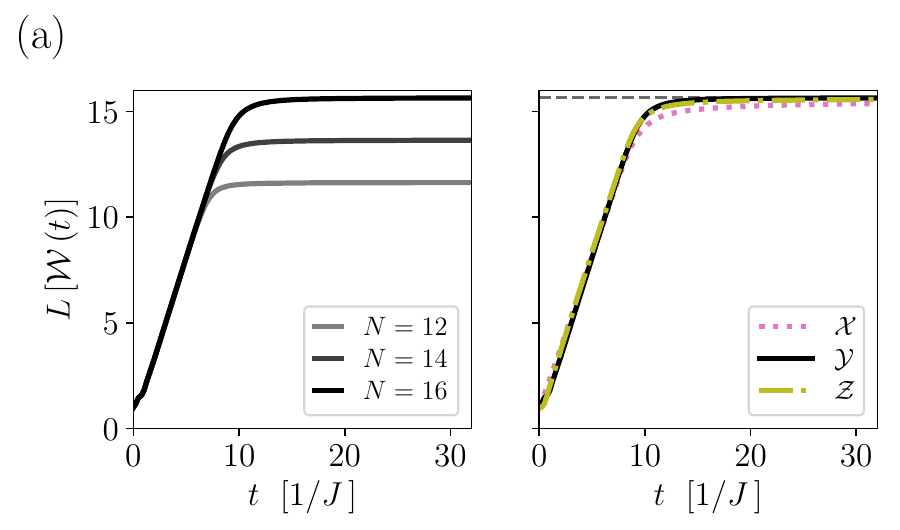}
    \includegraphics[width=.98\linewidth]{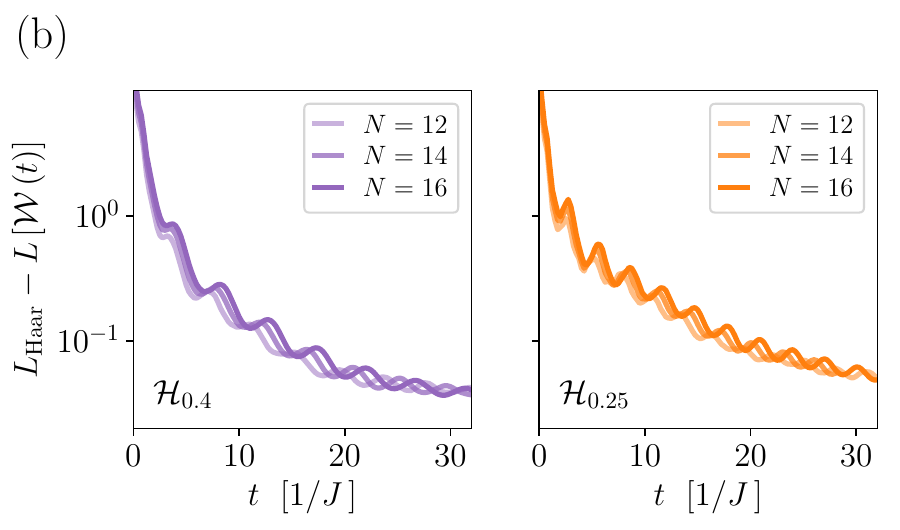}
    \caption{(a) Temporal evolution of the operator size for the local Hamiltonian $\mathcal{H}_\infty$. Left panel shows the operator size for $\mathcal{W}=\mathcal{Y}$ and various system sizes $N$. Right panel shows the operator sizes for various choices of $\mathcal{W}$ and $N=16$. (b) Slowdown of operator size for small values of the decay exponent $\alpha$. The approach towards the expected late-time value $L_\mathrm{Haar}$ is displayed for $\mathcal{W}=\mathcal{Y}$ and different system sizes $N$. Left panel shows data for $\alpha=0.4$ and right panel for $\alpha=0.25$. Data is obtained using exact diagonalization (ED).} 
    \label{fig:op_size_app}
\end{figure}

This section provides additional data regarding the operator size, which is summarized in Fig.\,\ref{fig:op_size_app}. In the left panel of Fig.\,\ref{fig:op_size_app}\,(a), we display the linear growth of the operator size regarding the local Hamiltonian $\mathcal{H}_\infty$, $\mathcal{W}=\mathcal{Y}$, and different system sizes $N$. All shown system sizes are characterized by the same linear growth of operator size. Moreover, the saturation value agrees with the expected late-time value $L_\mathrm{Haar}$ for the respective system size. The right panel of Fig.\,\ref{fig:op_size_app}\,(a), shows the operator size regarding the local Hamiltonian $\mathcal{H}_\infty$ for $N=16$ and different choices of $\mathcal{W}$. Although the growth of the operator size is similar in all cases, it appears that for $\mathcal{W}=\mathcal{X},\mathcal{Z}$ the saturation value is slightly smaller than $L_\mathrm{Haar}$, which we attribute to the non vanishing overlap between these local operators and the Hamiltonian \cite{altman_op_universival}.   

Further data regarding the slow approach of the operator size towards its late-time value for small decay exponents $\alpha$ is presented in Fig.\,\ref{fig:op_size_app}\,(b). Accordingly, we display $L_\mathrm{Haar}-L\left[\mathcal{W}\left(t\right)\right]$ for $\mathcal{W}=\mathcal{Y}$ and different system sizes $N$. The left panel is associated with $\alpha=0.4$ and the right panel with $\alpha=0.25$. In both cases, the slowdown of operator size is robust upon increasing the system size $N$.

\section{Convergence of TDVP results}
\label{app:tdvp}

In this section we provide additional details regarding our computations using matrix product states. All the results have been obtained using a single site time-dependent variational principle (TDVP) update \cite{tdvp}. For the calculation of the operator density we have used a state representation of the operator \cite{lr_lightcone_levy_swingle} defined in a doubled Hilbert space. 

\begin{figure}[h]
    \centering
    \includegraphics[width=.98\linewidth]{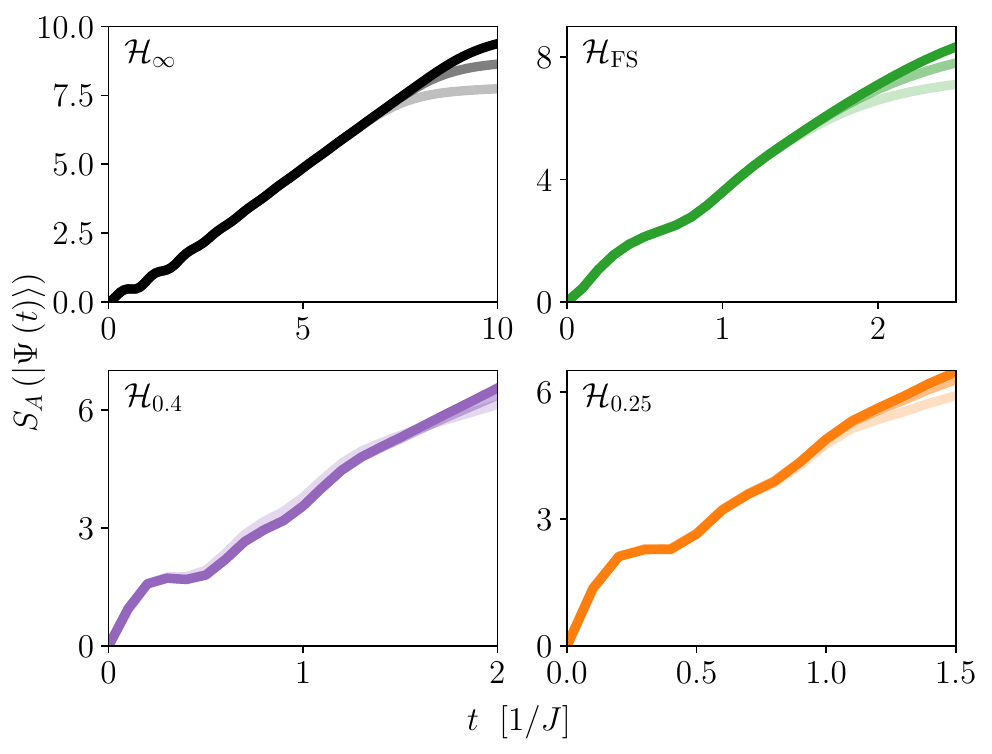}
    \caption{Half-chain entanglement entropy for $N=32$ and various bond dimensions $\chi=\left\{256,512,1024\right\}$ (TDVP). Time step of the simulation is $\delta t=0.1 \left[1/J\right]$. Darker colors indicate larger bond dimension. Local Hamiltonian $\mathcal{H}_\infty$ (upper left), fast scrambler $\mathcal{H}_\mathrm{FS}$ (upper right), power law Hamiltonian $\mathcal{H}_\alpha$ with $\alpha=0.4$ (lower left) and $\alpha=0.25$ (lower right).} 
    \label{fig:tdvp_conv}
\end{figure}

\begin{figure}[h]
    \centering
    \includegraphics[width=.98\linewidth]{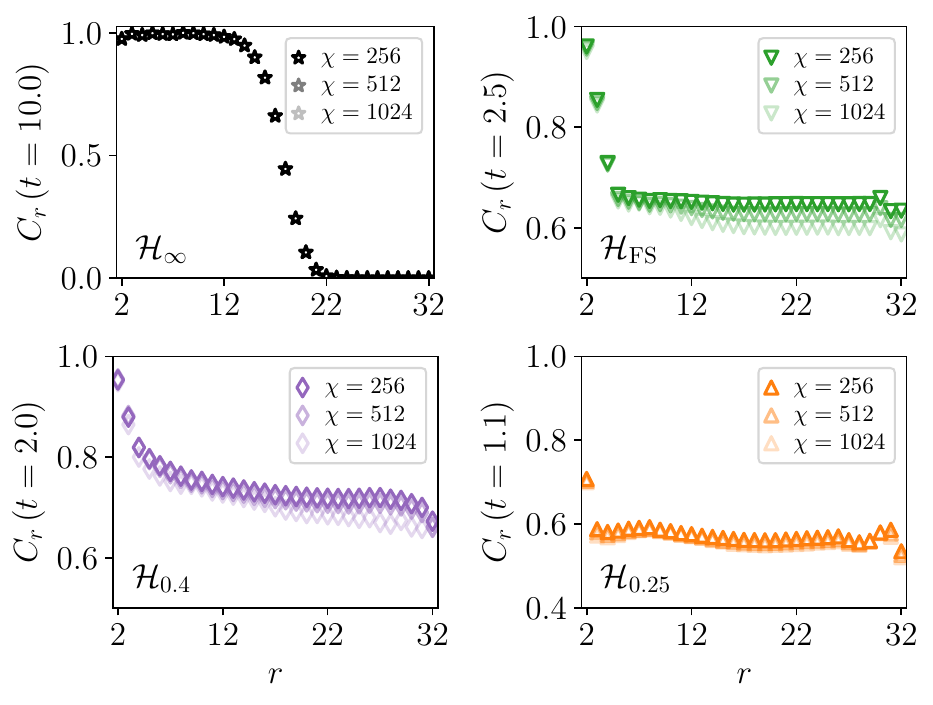}
    \caption{Squared commutator evaluated in the initial state $\ket{Y+}$ at a fixed time $t$ for $\mathcal{W},\mathcal{V}=\mathcal{Y}$, $N=32$, and various bond dimensions $\chi=\left\{256,512,1024\right\}$ (TDVP). Time step of simulation is $\delta t=0.1 \left[1/J\right]$. Local Hamiltonian $\mathcal{H}_\infty$, $t=10.0$ (upper left), fast scrambler $\mathcal{H}_\mathrm{FS}$, $t=2.5$ (upper right), power law Hamiltonian $\mathcal{H}_\alpha$ with $\alpha=0.4$, $t=2.0$ (lower left), and $\alpha=0.25$, $t=1.5$ (lower right).} 
    \label{fig:tdvp_conv2}
\end{figure}

We have performed convergence checks of the quantities of interest with increasing bond dimension $\chi$. In Fig.\,\ref{fig:tdvp_conv}, the half-chain entanglement entropy following a quench with initial state $\Ket{Y+}$ is shown for various bond dimensions. For the time intervals we considered, the entanglement entropy is clearly converged. Only small deviations at the end of the respective time intervals can be observed. The squared commutator seems to be more sensitive and a larger bond dimension is needed for convergence, see Fig.\,\ref{fig:tdvp_conv2}, as one has to calculate forward and backward evolution, requiring effectively a simulation of twice the time-scales.

\begin{figure}[h]
    \centering
    \includegraphics[width=.98\linewidth]{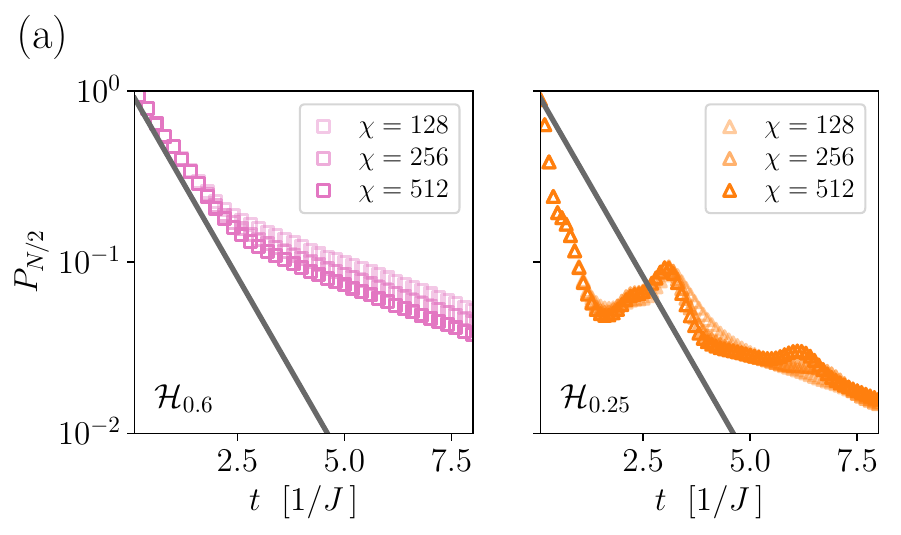}
    \includegraphics[width=.98\linewidth]{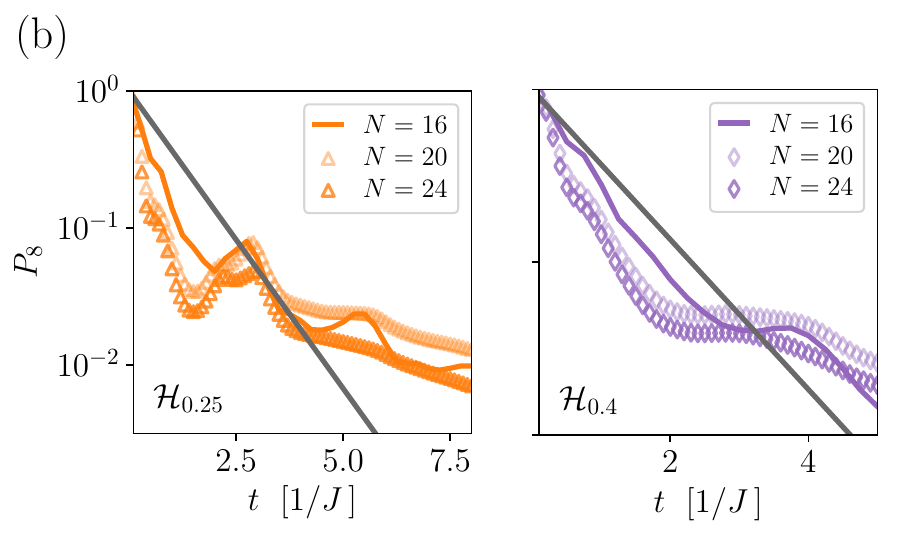}
    \caption{(a) Slow decay of operator density in the left half of the system for $\mathcal{W}=\mathcal{Y}$, various bond dimensions $\chi=\left\{256,512,1024\right\}$, and a system size of $N=24$ (TDVP). Left panel shows data for $\alpha=0.6$ and right panel for $\alpha=0.25$. (b) Operator density on the first eight sites for $N=16,20,24$ respectively. Left panel is data for $\alpha=0.25$ and right panel for $\alpha=0.4$.} 
    \label{fig:tdvp_conv3}
\end{figure}

In Fig.\,\ref{fig:tdvp_conv3}, we display additional data regarding the operator density. As discussed in the main text, we observe a slowdown in the decay of the operator density also for other values of $\alpha$, which is shown in Fig.\,\ref{fig:tdvp_conv3}\,(a) for $\alpha=0.25$ and $\alpha=0.6$ respectively. Also for these values, the slowdown remains upon increasing bond dimension. Furthermore, this slowdown is robust against increasing the system size $N$, which is depicted in Fig.\,\ref{fig:tdvp_conv3}\,(b) for $\alpha=0.25$ and $\alpha=0.4$ respectively.

\section{Operator Density and the Squared Commutator}
\label{app:op_dens_alg}
In the following, we present more details on the relationship between the operator density and the squared commutator. To this end, let us consider the squared commutator, where the operator $\mathcal{W}$ is initially placed at the left edge of the system, and the operator $\mathcal{V}$ at site $r$ 
\begin{align}
\label{otoc_app}
    C^\mathcal{V}_r\left(t\right)=\frac{1}{2}\left\langle\left[\mathcal{W}\left(t\right),\mathcal{V}_r\right]^\dagger\left[\mathcal{W}\left(t\right),\mathcal{V}_r\right]\right\rangle\,.
\end{align}
Here, we use $\left\langle\ldots\right\rangle=2^{-N}\Tr\left(\ldots\right)$, which is the expectation value in the infinite temperature ensemble. Furthermore, we assume $\mathcal{W}$ and $\mathcal{V}$ to be unitary. Let us first consider $r=N$, Eq.\,\eqref{otoc_app} then reads
\begin{align}
    \label{otoc_app2}
    &\frac{1}{2}\left\langle\left[\mathcal{W}\left(t\right),\mathcal{V}_N\right]^\dagger\left[\mathcal{W}\left(t\right),\mathcal{V}_N\right]\right\rangle\nonumber\\
    &=\left\langle\left[\mathbb{P}_N\mathcal{W}\left(t\right)\right]^2\right\rangle-\left\langle\left[\mathbb{P}_N\mathcal{W}\left(t\right)\mathcal{V}_N\right]^2\right\rangle\,,
\end{align}
where we defined $\mathbb{P}_N\mathcal{W}\left(t\right):=\sum_{\left\vert\Lambda\right\vert=N}c_\Lambda\mathcal{S}_\Lambda$ as the projection of $\mathcal{W}\left(t\right)$ onto strings that act nontrivially on site $N$. The first term in Eq.\,\ref{otoc_app2} then reads
\begin{align}
\label{otoc_app_term1}
\left\langle\left[\mathbb{P}_N\mathcal{W}\left(t\right)\right]^2\right\rangle&=\sum_{\left\vert\Lambda\right\vert=\left\vert\Lambda^\prime\right\vert=N}c^*_\Lambda c_{\Lambda^\prime} \left\langle\mathcal{S}_\Lambda \mathcal{S}_{\Lambda^\prime}\right\rangle\nonumber\\
&=\sum_{\left\vert\Lambda\right\vert=N}\left\vert c_\Lambda\right\vert^2=p_N\left(t\right)\,,
\end{align}
which is just the operator density for $r=N$. For the second term, we obtain
\begin{align}
\label{otoc_app_term2}
    \left\langle\left[\mathbb{P}_N\mathcal{W}\left(t\right)\mathcal{V}_N\right]^2\right\rangle&=\sum_{\left\vert\Lambda\right\vert=\left\vert\Lambda^\prime\right\vert=N}c^*_\Lambda c_{\Lambda^\prime}\left\langle\mathcal{S}_{\Lambda}\mathcal{V}_N\mathcal{S}_{\Lambda^\prime}\mathcal{V}_N\right\rangle\nonumber\\
    &=\sum_{\left\vert\Lambda\right\vert=N, \Lambda_N=\mathcal{V}}\left\vert c_\Lambda \right\vert^2-\sum_{\left\vert\Lambda\right\vert=N, \Lambda_N\neq\mathcal{V}}\left\vert c_\Lambda \right\vert^2\,.
\end{align}

The second line follows from the fact that $\mathcal{V}_N\mathcal{S}_{\Lambda}\mathcal{V}_N=\pm\mathcal{S}_{\Lambda}$, where we obtain a negative sign if the string $\mathcal{S}_\Lambda$ at site $N$ is not $\mathcal{V}$. Combining Eq.\,\eqref{otoc_app_term1} and \eqref{otoc_app_term2} we obtain
\begin{align}
\label{otoc_app3}
    C_N^\mathcal{V}\left(t\right)=2\sum_{\left\vert\Lambda\right\vert=N, \Lambda_N\neq\mathcal{V}}\left\vert c_\Lambda \right\vert^2\,.
\end{align}
Note that Eq.\,\eqref{otoc_app} depends on the choice of $\mathcal{V}$. We can define an average square commutator as 
\begin{align}
    \label{otoc_average}
    \overline{C}_r\left(t\right)=\frac{1}{\left\vert\mathcal{P}\right\vert}\sum_{\mathcal{V}\in\mathcal{P}}C_r^\mathcal{V}\left(t\right)=\sum_{\Lambda_r\neq\mathbf{1}}\left\vert c_\Lambda\right\vert^2\,,
\end{align}
where $\mathcal{P}=\left\{\mathbf{1},\mathcal{X},\mathcal{Y},\mathcal{Z}\right\}$. Hence, we can establish the following equality between the squared commutator and the operator density
\begin{align}
\label{otoc_dens_N}
    p_N\left(t\right)=\overline{C}_N\left(t\right)\,.
\end{align}
Equation\,\eqref{otoc_dens_N} is a special case. In general, the average squared commutator $\overline{C}_r\left(t\right)$ is determined by all coefficients $c_\Lambda$ that belong to strings $\mathcal{S}_\Lambda$ that act nontrivially on site $r$. For $r=N$, this coincides with all coefficients that belong to strings of size $N$. In the general case, we obtain

\begin{align}
    p_r\left(t\right)=\overline{C}_r\left(t\right)-\sum_{\left\vert\Lambda\right\vert>r, \Lambda_r\neq\mathbf{1}}\left\vert c_\Lambda \right\vert^2\,.
\end{align}

Thus, in this general case, the operator density is bounded from above by the average squared commutator, i.e., 
\begin{align}
    p_r\left(t\right)\leq\overline{C}_r\left(t\right)\,,\, r>1\,.
\end{align}

If one has access to the Heisenberg operator $\mathcal{W}\left(t\right)$, for instance, within an ED computation, the operator density can be obtained as follows: 
the part of the operator whose support ranges up to a given site $\ell$ can be obtained by taking the partial trace with respect to all sites to the right of $\ell$, i.e.,
\begin{align}
    \mathcal{W}_{\ell}\left(t\right):=\frac{1}{2^{N-\ell}}\Tr_{\overline{\mathcal{L}}}\left[\mathcal{W}\left(t\right)\right]=\sum_{ \left\vert\Lambda\right\vert\leq\ell}c_\Lambda\left(t\right)\mathcal{S}_\Lambda\,,
\end{align}
where $\overline{\mathcal{L}}$ is the complement of $\left\{1,\ldots,\ell\right\}$. Note that $\mathcal{W}_{\ell}\left(t\right)$ is not unitary anymore. It follows then straightforwardly that

\begin{align}
    \label{op_dens_calc}
    \left\langle \mathcal{W}_{\ell}\left(t\right) \mathcal{W}_{\ell}\left(t\right)\right\rangle=\sum_{\ell^\prime\leq\ell}p_{\ell^\prime}\left(t\right)\,.
\end{align}

Thus, by computing Eq.\,\eqref{op_dens_calc} for all $1\leq\ell\leq N-1$, one can reconstruct the operator density for all $\ell$. In particular, we have
\begin{align}
    p_\ell\left(t\right)=\left\langle\mathcal{W}_\ell\left(t\right)\mathcal{W}_\ell\left(t\right)\right\rangle - \left\langle\mathcal{W}_{\ell-1}\left(t\right)\mathcal{W}_{\ell-1}\left(t\right)\right\rangle\,.
\end{align}

\bibliography{ref}

\end{document}